\documentclass[acmsmall,authorversion,timestamp,nonacm,screen]{acmart}

\acmJournal{PACMPL}
\acmVolume{1}
\acmNumber{CONF} 
\acmArticle{1}
\acmYear{2018}
\acmMonth{1}
\acmDOI{} 
\startPage{1}

\setcopyright{none}

\usepackage{booktabs}   
\usepackage{subcaption} 
\usepackage{custom}     
\usepackage{notation}   
\usepackage{pgf}
\usepackage{amsmath}
\usepackage{textcomp}
\usepackage{algpseudocode}

\begin{document}

\title{Lay-it-out: Interactive Design of Layout-Sensitive Grammars}
\subtitle{(Preliminary Draft)}          

\author{Fengmin Zhu}
\authornote{Both authors contributed equally to this work.}
\authornote{Early revisions of this work were done when the author was in Tsinghua University.}          
\affiliation{
  \institution{MPI-SWS}            
  \city{Saarbrücken}
  \state{Saarland}
  \postcode{66123}
  \country{Germany}                    
}
\email{paulzhu@mpi-sws.org}          
\affiliation{
  \institution{Tsinghua University}       
  \city{Beijing}
  \postcode{100084}
  \country{China}                                
}

\author{Jiangyi Liu}
\authornotemark[1]
\affiliation{
  \institution{Tsinghua University}       
  \city{Beijing}
  \postcode{100084}
  \country{China}                                
}
\email{liujiang19@mails.tsinghua.edu.cn}          

\author{Fei He}
\affiliation{
  \institution{Tsinghua University}       
  \city{Beijing}
  \postcode{100084}
  \country{China}                     
}
\email{hefei@tsinghua.edu.cn}          

\begin{abstract}
  Layout-sensitive grammars have been adopted in many modern programming languages.
However, tool support for this kind of grammars still remains limited and immature.
In this paper, we present \ourtool, an \emph{interactive} framework for layout-sensitive grammar design.
Beginning with a user-defined ambiguous grammar, our framework refines it by synthesizing \emph{layout constraints} through user interaction.
For ease of interaction, a \emph{shortest} nonempty ambiguous sentence (if exists) is automatically generated by our \emph{bounded ambiguity checker} via SMT solving.
The \emph{soundness} and \emph{completeness} of our SMT encoding are mechanized in the Coq proof assistant.
Case studies on real grammars, including a \emph{full grammar}, demonstrate the practicality and scalability of our approach.

\end{abstract}



\maketitle

\section{Introduction}

\emph{Layout-sensitive} (or \emph{indentation-sensitive}) grammars were first proposed by \citet{Landin}.
Nowadays, they have been adopted in many programming languages, e.g.
Python \cite{Python}, Haskell \cite{Haskell-10}, F\# \cite{FSharp}, Yaml \cite{Yaml} and Markdown \cite{Markdown}.
These grammars are also introduced as new features in latest versions of languages, e.g. optional braces in Scala 3\footnote{
    \url{https://docs.scala-lang.org/scala3/reference/other-new-features/indentation.html}
}.
Due to the presence of layout constraints, indentations and whitespaces affect how a program should be parsed.
Although this amazing feature gives rise to a stylized, structural and elegant syntax,
it \emph{increases the complexity} of theoretical study (e.g. ambiguity problem) and tool development.

In the recent decade, pioneering studies have been made on \emph{layout-sensitive parsing}
\cite{Revisiting-Offside,Layout-GLR,Decl-Spec}.
To apply these parsing techniques, one must, first of all, have the grammar formally specified.
This grammar, in a serious language design phase, should be \emph{unambiguous}.
Due to the complexity introduced by layout constraints,
\emph{manually} checking the ambiguity (especially for large grammars) is \emph{tedious} and \emph{error-prone}.
Therefore, tool support for \emph{automated} ambiguity checking is essential to \emph{reduce human labor} and \emph{avoid human mistakes}.

For \emph{context-free grammars} (CFGs), a common practical approach of ensuring unambiguity
is to check if it belongs to an \emph{unambiguous fragment} of CFG, such as $LL(k)$ \cite{DragonBook} and $LR(k$) \cite{LRk}.
Off-the-shelf tools such as Lex/Yacc \cite{Lex-Yacc}, Flex/Bison \cite{Flex-Bison}, and SDF \cite{SDF} have been built to automatically check whether a user-input CFG is $LL(k)$ or $LR(k)$ (and generate the parser as well).
For layout-sensitive grammars, however, this approach is not yet practical: as far as we know, there is no similar theory or tool support.

In grammar design, proving unambiguity is an ultimate goal but it is \emph{hard}:
checking the ambiguity of a CFG is already \emph{undecidable} \cite{Algebraic-Theory-CFL,ATLC},
so it is even harder for a layout-sensitive grammar (more expressive than CFG).
Alternatively, we aim to tackle the \emph{bounded ambiguity problem},
which is more restrictive and thus decidable, but yet still \emph{practical}:
is there any ambiguous sentence (\ie has at least two different parse trees) within a given length of a grammar?
A sound and complete checker for this problem can \emph{automatically} detect ambiguity
during the entire process of grammar design, which helps to \emph{exposes human mistakes} as early as possible.

In this paper, we present a novel interactive framework, \ourtool, for layout-sensitive grammar design.
This framework saves time and efforts of grammar designers in two aspects:
(1) a \emph{bounded ambiguity checker} automatically generates a \emph{shortest} nonempty ambiguous sentence (if any) so that the grammar designer can detect the cause of ambiguity by a \emph{simplest} and \emph{concrete} example;
(2) a \emph{layout constraints synthesizer} recommends candidate \emph{layout constraints} to refine the input-grammar via user interaction based on the ambiguous sentence,
so that the user needs \emph{not manually} specify any layout rules.
This feature also helps an \emph{ordinary developer} who knows little about how to specify layout rules, but wants to design a layout-sensitive grammar for their own domain-specific languages.

\begin{figure}
    \centering
    \includegraphics[width=.6\textwidth]{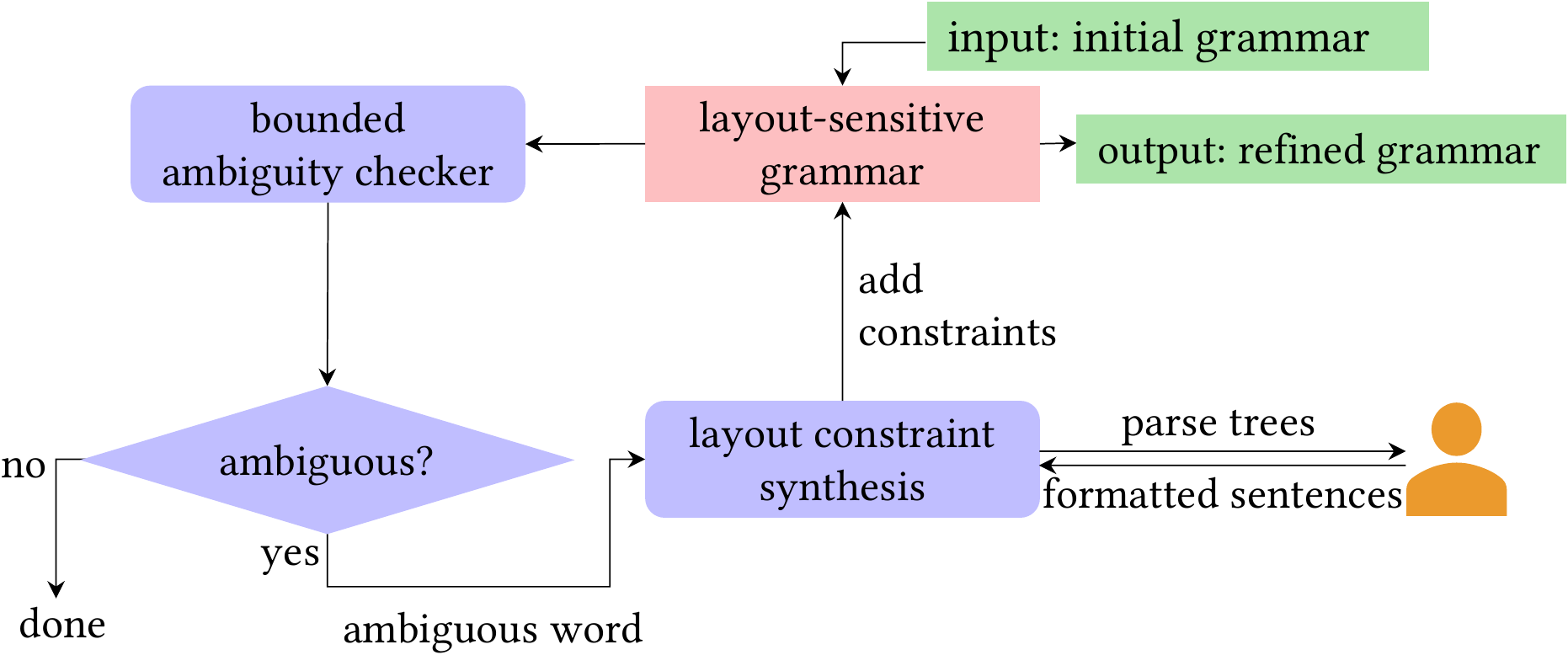}
    \caption{High-level framework of \ourtool.}
    \label{fig:framework}
\end{figure}

In a nutshell, the workflow of \ourtool is a grammar \emph{refinement} loop
guided by the detection of ambiguous sentences, as depicted in \cref{fig:framework}.
In the initial round, the user specifies a layout-free grammar (\ie a CFG) as input.
Then, the bounded ambiguity checker produces a shortest nonempty ambiguous sentence (if any) and present it with all its parse trees to the user.
Due to inadequate layout constraints in the input grammar, these parse trees shall be distinguishable once more layout constraints are added.
The missing layout constraints are synthesized from user's \emph{feedback} thats reflect his/her intents:
the user is asked to \emph{format} (e.g. insert indentations and newlines between tokens) the ambiguous sentence in distinct ways to match each parse tree.
The synthesizer recommends a set of candidate layout constraints: the user can review and select the ones that meet his/her intents.
The user-selected layout constraints will be added to the input grammar.
Multiple interaction rounds may be taken until the user is satisfied with the refined grammar.

SMT (\emph{Satisfiability Modulo Theories}) solving is a powerful technique to
verification \cite{SMT-Why-How,SMT-Program-Verification,Unifying-View,Dafny} and
synthesis \cite{Oracle-Guided,Superoptimization,Quantifier-Instantiation}.
We find it also suitable for building a bounded ambiguity checker.
The layout constraints contain relational operations (e.g. $<$, $>$, $=$) on line/column numbers of tokens -- they can be easily expressed in \emph{integer difference logic} that is supported by most SMT solvers.

To apply SMT solving, one must encode the problem as a SMT formula.
Instead of directly using the standard notion of ambiguity, we propose an alternative definition called \emph{local ambiguity}, which leads to a more \emph{straightforward} and \emph{efficient} encoding.
The skeleton of our local ambiguity is inspired by \tool{CFGAnalyzer} \cite{CFG-SAT}, a tool that studies meta properties of CFGs, including bounded ambiguity, via SAT solving.
Our key innovation lies in a \emph{reachability} relation defined for layout-sensitive grammars, taking the influence of layout constraints on ambiguity into careful account.
In this way, we can show the \emph{logical equivalence} between local ambiguity and standard ambiguity.

With the equivalence theorem, we are able to construct a \emph{sound} and \emph{complete} SMT encoding for solving the bounded ambiguity problem on \emph{acyclic} layout-sensitive grammars.
Being acyclic is \emph{not} a strong assumption: well-designed grammars should have this nice property.
The definitions and theorems related to local ambiguity and encoding are mechanized in the Coq proof assistant.
In fact, Coq helped us a lot in finding pitfalls on early revisions of the encoding.

Our layout constraint synthesizer performs analysis on the position information conveyed in user-provided formatted words, and computes a set of candidate layout constraints that \emph{refine} the input grammar.
We allow the user to accept a subset of the candidate based on his/her intents.
Note that our synthesizer is targeting the disambiguation caused by inadequate layout constraints, which is typical and unique for layout-sensitive grammars.
Other kinds of ambiguity such as the well-known binary operator precedence (were studied in CFGs \cite{Parsify,Disambig-Filters,Disambig-Sub-parse}) are not focused on in this work.

To measure the practicality of our approach, we conducted case studies on real grammars in different scales (the initial inputs are their layout-free versions):
(1) 6 representative grammar fragments extracted from Python, F\# and Haskell (small-scale);
(2) the layout-sensitive subset of YAML, a markup language (middle-scale);
(3) the \emph{full grammar} of SASS, a styling language for the web (large-scale).
For the last case, 74 layout constraints were introduced in 16 interaction rounds,
with the longest ambiguous sentence of length 12.

To sum up, this paper makes the following contributions:
\begin{itemize}
    \item \ourtool, the first (to the best of our knowledge) interactive framework for layout-sensitive grammar design (\cref{sec:example}, \cref{sec:synthesis}, \cref{sec:implementation});
    \item local ambiguity, an equivalent ambiguity definition that is SMT-friendly (\cref{sec:ambiguity});
    \item bounded ambiguity checker, developed from a sound and complete SMT encoding for acyclic grammars (\cref{sec:encoding}, \cref{sec:coq});
    \item case studies on real grammars of different scales, including a full grammar, shows the practicality and scalability of \ourtool (\cref{sec:evaluation}).
\end{itemize}

\section{Preliminary}\label{sec:grammar}

In formal language theory, a grammar $G$ is a quadruple $(N,\Sigma,P,S)$,
where $N$ is a finite (nonempty) set of \emph{nonterminals},
$\Sigma$ is a finite set of \emph{terminals} (or \emph{tokens}),
$P\subseteq N\times(N\cup\Sigma)^{*}$ is a finite set of \emph{production rules},
and $S \in N$ is the \emph{start symbol}.

In a CFG, a \emph{sentence} (sometimes also called a \emph{word}) is a token sequence.
In a \emph{layout-sensitive grammar}, a sentence is a sequence of \emph{positioned tokens}\footnote{
    We assume all sentences are well-formed: positions of tokens must be in ascending order.
    Ill-formed sentences do not physically exist in reality,
    e.g. $[a @ (1,2), b @ (1,1)]$ ($b$ comes before $a$) and $[a @ (1,1), b @ (1,1)]$ ($a$ and $b$ overlap).
}.
Each positioned token has the form $a @ (line, col)$, where $a \in \Sigma$ gives the terminal,
and $(line, col)$ gives the position (i.e. line and column number) in the source file.
We denote an empty sentence by $\emptystr$.
For a nonempty sentence $w$, we denote its length by $|w|$.
We write $w[i]$ to access the positioned token at index $i$, and $\Term{w[i]}$, $\Line{w[i]}$ and $\Col{w[i]}$ resp. the terminal, line number and column number.

\paragraph{Layout Constraints}\label{sec:grammar:constraints}

An unary (resp. binary) \emph{layout constraint} $\varphi$ is a logical predicate over one (resp. two) sentence(s),
specifying positional restrictions on tokens of the sentences.
As a convention, no constraint can be given to empty sentences:
$\varphi(\emptystr) = \true$ (resp. $\varphi(\emptystr, \cdot) = \varphi(\cdot, \emptystr) = \true$ for binary case).

We support the following built-in layout constraints: alignment (binary), indentation (binary), offside (unary), offside-align (a variant of offside, unary) and single-line (unary).
The first three are widely-used in many practical grammars, and they have been well-studied in previous work \cite{Decl-Spec}.
Offside-align and single-line rules are what we find useful in case studies (\cref{sec:evaluation}).
Their logical predicates are presented in \cref{fig:layout-predicate},
where $\tl{w}$ denotes the last positioned token of $w$.

\begin{figure*}
    \small
    \begin{mathpar}
        \thely{align}(w_1, w_2) \defas (w_1 \neq\emptystr \land w_2 \neq\emptystr)
             \implies \Col{\hd{w_1}} = \Col{\hd{w_2}}
        \and \thely{indent}(w_1, w_2) \defas (w_1 \neq\emptystr \land w_2 \neq\emptystr)
             \implies (\Col{\hd{w_2}} > \Col{\hd{w_1}} \land \Line{\hd{w_2}} = \Line{\tl{w_1}} + 1)
        \and \thely{offside}(w) \defas w \neq\emptystr \implies
             \forall t \in w~\text{s.t.}~\Line{t} > \Line{\hd{w}}, \Col{t} > \Col{\hd{w}}
        \and \thely{single}(w) \defas w \neq\emptystr \implies
             \forall t \in w, \Line{t} = \Line{\hd{w}}
    \end{mathpar}
    \caption{Logical predicates of built-in layout constraints.}
    \label{fig:layout-predicate}
\end{figure*}

Alignment restricts the first token of the two sentences to be aligned at column.
Indentation restricts the second part has its first token to the right (at column) of the first part,
and a newline exists in between.
The offside rule was first proposed by \citet{Landin} and later revised by \citet{Revisiting-Offside}.
It restricts that in the parsed sentence,
any subsequent lines must start from a column that is further to the right of the start token of the first line.
We also consider offside-align rule, a variant of the above,
where subsequent lines can start from the same column as that of the first line.
The single-line rule restrict the parsed sentence to be one-line.

\paragraph{Binary Normal Form}
It is well-known~\cite{lange2009cnf} that every CFG can be converted to an equivalent \emph{binary normal form} (or \emph{Chomsky normal form}).
We migrate the binary normal form to layout-sensitive grammars as follows.

\begin{definition}
A layout-sensitive grammar is in \emph{binary normal form}, called LS2NF,
if for every production rule, its right-hand side (called a \emph{clause}) is
one of the following:
\begin{itemize}
     \item an empty clause $\varepsilon$;
     \item an atomic clause $a$, where $a \in \Sigma$;
     \item an unary clause $A^\varphi$, where $A \in N$ and $\varphi$ is a \emph{unary layout constraint};
     \item a binary clause $A \mathbin\varphi B$, where $A, B \in N$ and $\varphi$ is a \emph{binary layout constraint}.
\end{itemize}
\end{definition}

Our definition is the same with CFG, but with layout constraints in unary/binary clauses.
To avoid inconsistent layout constraints, the production rule set cannot contain the following rules simultaneously:
(1) $A \to B^\varphi$ and $A \to B^{\varphi'}$ where $\varphi \neq \varphi'$;
(2) $A \to B_1 \mathbin\varphi B_2$ and $A \to B_1 \mathbin\varphi' B_2$ where $\varphi \neq \varphi'$.

Throughout the paper, we study layout-sensitive grammars in LS2NF,
and use \emph{Extended Backus-Naur form} (EBNF) to express complex grammars for readability:
they are converted to equivalent LS2NF in the standard way \cite{lange2009cnf}.
We also use notations in place of rule names for built-in layout constraints:
we write $\alpha \alignto \beta$ for $\alpha \ly{align} \beta$,
$\alpha \indents \beta$ for $\alpha \ly{indent} \beta$,
$\offside\alpha$ for $\alpha^{\ly{offside}}$,
$\offsidealign{\alpha}$ for $\alpha^{\ly{offside-align}}$
and $\oneline\alpha$ for $\alpha^{\ly{single}}$.
The alignment constraint is extended to lists in EBNF: $\alignplus{\alpha}$ and $\alignstar{\alpha}$ denote all sentences parsed by $\alpha$ are aligned.

\paragraph{Parse Trees}

\emph{Parse trees} are rooted trees (let $A \in N$ be the root) inductively defined as follows:
\begin{align*}
     t &::= \emptytree{A} \mid \tokentree{A}{a @ (i, j)} \mid \unarytree{A}{t} \mid \binarytree{A}{t_1}{t_2}
\end{align*}
They resp. correspond to the derivation using the empty, atom, unary and binary clause.
We write $\rootof t$ and $\wordof t$ to denote the root node and the represented sentence of a parse tree $t$.
A parse tree is said \emph{valid} if it follows the production rules and fulfills the layout-constraints:
\begin{itemize}
    \item $\emptytree{A}$ is valid if $A \to \varepsilon \in P$;
    \item $\tokentree{A}{a @ (i, j)}$ is valid if $A \to a \in P$;
    \item $\unarytree{A}{t}$ is valid if $A \to \rootof{t}^\varphi \in P$, $\varphi(\wordof{t})$ is true
        and $t$ is valid;
    \item $\binarytree{A}{t_1}{t_2}$ is valid if $A \to \rootof{t_1} \mathbin\varphi \rootof{t_2} \in P$,
        $\varphi(\wordof{t_1}, \wordof{t_2})$ is true and both $t_1$, $t_2$ are valid.
\end{itemize}

We say a parse tree $t$ \emph{witnesses} a derivation from a nonterminal $A$ to a sentence $w$,
written $t \witness A \derive w$, if $t$ is valid, $\rootof t = A$ and $\wordof t = w$.
Using the above notions, the usual notion of derivation $A \derive w$ is actually a short-hand for
``$\exists t, t \witness A \derive w$''.
Specially, we use the predicate $\nullable{A}$ to indicate that $A \derive \emptystr$, a.k.a. $A$ is \emph{nullable}.

\paragraph{Derivation Ambiguity}

We say the derivation $A \derive w$ is \emph{ambiguous} if there exist at least two different parse trees that witness $A \derive w$.
In this way, the bounded ambiguity problem is restated as:
for a given bound $k$, is there a $w$ s.t. $|w| \le k$ and $S \derive w$ is ambiguous?

Throughout the paper, we reserve $A, B, C$ for nonterminals,
$a, b, c$ for tokens/terminals, $w$ for sentences, $t, T$ for parse trees,
and $\varphi$ for both unary and binary layout constraint (inferred from context).

\section{A Running Example}\label{sec:example}

In this section, we illustrate the workflow of \ourtool with a toy grammar $G_\nt{block}$:
\begin{align*}
    \nt{block} &\to \nt{stmt}^{+} \\
    \nt{stmt} &\to \token{nop} \mid \token{do}~\nt{block}
\end{align*}
Intuitively, a block is a sequence of statements, and each statement is either a \token{nop} (no operation),
or a \token{do}-block whose body is recursively a block.
Apparently, $G_\nt{block}$ is ambiguous.

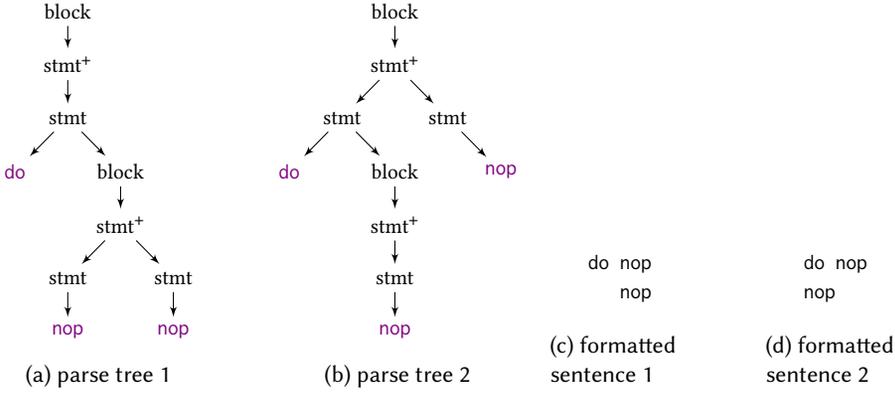
\begin{figure}
    \centering
    \centering
\begin{minipage}[b]{.28\textwidth}
\centering\footnotesize
\vspace{0pt}
\begin{tikzpicture}[draw, -latex', level distance=20, sibling distance=40]
    \node {$\nt{block}$}
        child {
            node {$\nt{stmt}^{+}$}
            child {
                node {$\nt{stmt}$}
                child {
                    node {$\token{do}$}  
                }
                child {
                    node {$\nt{block}$}
                    child {
                        node {$\nt{stmt}^{+}$}
                        child {
                            node {$\nt{stmt}$}
                            child {
                                node {$\token{nop}$}  
                            }
                        }
                        child {
                            node {$\nt{stmt}$}
                            child {
                                node {$\token{nop}$}  
                            }
                        }
                    }
                }
            }
        }
    ;
\end{tikzpicture}
\subcaption{parse tree 1}
\label{fig:tree-1}
\end{minipage}
\begin{minipage}[b]{.28\textwidth}
\centering\footnotesize
\vspace{0pt}
\begin{tikzpicture}[draw, -latex', level distance=20, sibling distance=40]
    \node {$\nt{block}$}
        child {
            node {$\nt{stmt}^{+}$}
            child {
                node {$\nt{stmt}$}
                child {
                    node {$\token{do}$}  
                }
                child {
                    node {$\nt{block}$}
                    child {
                        node {$\nt{stmt}^{+}$}
                        child {
                            node {$\nt{stmt}$}
                            child {
                                node {$\token{nop}$} 
                            }
                        }
                    }
                }
            }
            child {
                node (a) {$\nt{stmt}$}
                child {
                    node [below right of = a] {$\token{nop}$}  
                }
            }
        }
    ;
\end{tikzpicture}
\subcaption{parse tree 2}
\label{fig:tree-2}
\end{minipage}
\begin{minipage}[b]{.2\textwidth}
\centering\footnotesize
\vspace{0pt}
\begin{lstlisting}
do nop
   nop
\end{lstlisting}
\subcaption{formatted \\ sentence 1}
\label{fig:formatted-1}
\end{minipage}
\begin{minipage}[b]{.2\textwidth}
\centering\footnotesize
\vspace{0pt}
\begin{lstlisting}
do nop
nop
\end{lstlisting}
\subcaption{formatted \\ sentence 2}
\label{fig:formatted-2}
\end{minipage}
    \caption{Parse trees of ``\t{do nop nop}''.}
    \label{fig:tree}
\end{figure}

\ourtool takes $G_\nt{block}$ as input, preprocesses it (desugar to LS2NF), and invokes the bounded ambiguity checker.
It produces a \emph{shortest} ambiguous sentence \t{do nop nop}.
We present this sentence with its parse trees (depicted in \cref{fig:tree}) to the user.
In parse tree 1, the last \token{nop} belongs to the \token{do} block.
In parse tree 2, the last \token{nop} belongs to the top-level block.
The user accepts both as reasonable parsings of the ambiguous sentence.
Layout constraints are needed to distinguish them.
The user decides to align the statements of one block, so that the last \token{nop} belongs to which block is clarified.
We request the user to intuitively provide this intent by formatting \t{do nop nop} to match the two parse trees:
\cref{fig:formatted-1} and \cref{fig:formatted-2}
for the parse tree depicted in \cref{fig:tree-1} and \cref{fig:tree-2} respectively.
\ourtool internally translates them into two positioned token sequences:
\begin{align*}
    w_1 &= [\t{do}@(1,1), \t{nop}@(1,4), \t{nop}@(2,4)] \\
    w_2 &= [\t{do}@(1,1), \t{nop}@(1,4), \t{nop}@(2,1)]
\end{align*}

Based on these formatted sentences, our layout constraint synthesizer generates two transformation rules:
$\{\nt{stmt}^{+} \leadsto \alignplus{\nt{stmt}}, \nt{stmt} \leadsto \offside{\nt{stmt}}\}$.
Applying them to $G_\nt{block}$ produces a refined grammar:
\begin{align*}
    \nt{block} &\to \alignplus{\offside{\nt{stmt}}} \\
    \nt{stmt} &\to \token{nop} \mid \token{do}~{\nt{block}}
\end{align*}
For this toy example, one interaction round is enough. In practice, multiple interaction rounds may needed to eventually disambiguate a grammar.

\section{Local Ambiguity}\label{sec:ambiguity}

Our grammar design framework is guided by ambiguous sentences generated via constraint solving.
The key technical problem is to find a SMT formula $\fAmb{k}$ such that it is \emph{satisfiable} iff there exists an ambiguous sentence of length $k$, and that every satisfying model corresponds to such an ambiguous sentence.
This problem is immediately solved once we construct a SMT encoding for derivation ambiguity: ``there exists two different parse trees that both witness a derivation''.
However, the \emph{difficulty} lies in how to encode the existential of two \emph{different} parse trees:
in general, the node pair that shows the difference can appear at an \emph{arbitrary} level (or depth) on the parse trees,
then enumerating every possibility yields a \emph{sophisticated} and \emph{inefficient} SMT encoding.

An \emph{indirect} but \emph{more efficient} approach is that, figure out an \emph{alternative definition} that is \emph{logically equivalent to} derivation ambiguity, while being \emph{conveniently encodable} in SMT theories.
This section will propose such a definition which we call \emph{local ambiguity}.
With local ambiguity, in the next section, we will be able to construct a \emph{sound} and \emph{complete} SMT encoding.

\subsection{Motivation: Lessons from a Failure Attempt}\label{sec:ambiguity:motivation}

\tool{CFGAnalyzer} \cite{CFG-SAT} is a previous work on bounded ambiguity checking of \emph{reduced} CFGs via SAT solving.
Their SAT encoding is based on a condition they call \emph{bAMB}\footnote{
    In their paper, ``bAMB'' stands for ``bounded ambiguity'', but that term has a different meaning in this paper.
}:
``there exists a nonterminal $A$ and a sentence $w$ such that $w$ has at least two different (valid) parse trees rooted at $A$ and differ in a node on level 1''.
Since parse trees are visualized representations of derivations, such two trees correspond to two different ways of deriving $w$, using
either two distinct production rules for $A$, or one rule in two distinct ways.
In terms of logical constraints, bAMB is \emph{weaker} than derivation ambiguity:
(1) $A$ is \emph{existentially} quantified but not given;
(2) the two parse trees differ in a node just below (\ie a child of) the root $A$, which is \emph{determined} and \emph{shallow}.

\begin{figure}
    \centering
    \tikzset{draw, -latex', level distance=24, sibling distance=30}
\begin{minipage}[b]{.3\linewidth}
    \centering
    $G_S:$
    \begin{mathpar}
        S \to A \alignto B \and A \to C \mid C' \\
        C \to a \and C' \to a \\
        B \to b
    \end{mathpar}
\end{minipage}
\begin{minipage}[b]{.2\linewidth}
    \centering
    \begin{tikzpicture}
        \node {$A$}
            child {
                node {$C$}
                child {
                    node {$a @ (1,3)$}
                }
            }
        ;
    \end{tikzpicture}

    \subcaption*{$t_1$ (valid)}
\end{minipage}
\begin{minipage}[b]{.2\linewidth}
    \centering
    \begin{tikzpicture}
        \node {$A$}
            child {
                node {$C'$}
                child {
                    node {$a @ (1,3)$}
                }
            }
        ;
    \end{tikzpicture}
    \subcaption*{$t_2$ (valid)}
\end{minipage}
\begin{minipage}[b]{.24\linewidth}
    \centering
    \begin{tikzpicture}
        \node {$S$}
            child {
                node {$t_i$}
            }
            child {
                node {$B$}
                child {
                    node {$b @ (2,4)$}
                }
            }
        ;
    \end{tikzpicture}
    \subcaption*{$T_i$ (invalid)}
\end{minipage}
    \caption{A counterexample where bAMB does not apply to $G_S$ (where $S$ is the start symbol).}
    \label{fig:counterexample}
\end{figure}
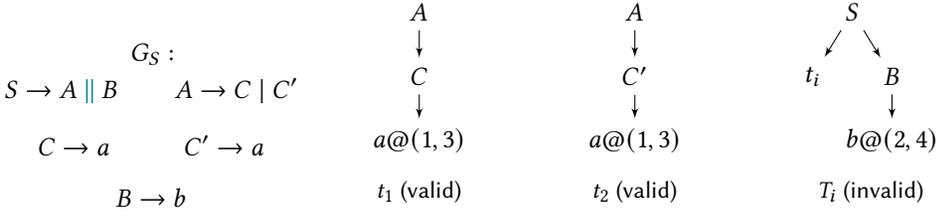

Our first attempt is to use bAMB as the alternative definition, so we must show ``bAMB $\Leftrightarrow$ derivation ambiguity'' for layout-sensitive grammars.
The $\Leftarrow$-part is trivial as we have seen that bAMB is weaker.
The $\Rightarrow$-part, however, does \emph{not} always hold, as demonstrated by \cref{fig:counterexample}:
According to the production rules of $G_S$, both $t_1$ and $t_2$ witness $A \derive [a @ (1,3)]$.
They differ in a node on level 1 ($C$ v.s. $C'$), which matches the definition of bAMB.
Since $A$ is not the start symbol, $t_1$ and $t_2$ alone cannot witness the ambiguity of $G_S$, \ie $S \derive w$ is ambiguous for some $w$.
To achieve that, we have to expand $t_1$ and $t_2$ each to a valid parse tree rooted at $S$ -- then the expanded trees must be distinct because $t_1 \neq t_2$.
Unfortunately, this is impossible:
$T_i$ ($i = 1,2$) are the only possible trees (having $t_i$ as a subtree) according to the production rules, but they are both \emph{invalid} because the \emph{alignment constraint} $A \alignto B$ is \emph{not fulfilled}, \ie $a @ (1,3)$ and $b @ (2,4)$ are at distinct columns ($3 \neq 4$).

We learn from the above example that the presence of layout constraints can affect parse tree validity in a \emph{non-obvious} way.
In \cref{fig:counterexample}, the invalidity of $T_i$ ($i = 1,2$) is caused by the nonfulfillment of the alignment constraint, which further prevents us from expanding $t_1$ and $t_2$ into two distinct valid parse trees rooted at the start symbol.
This never happens to CFGs: we can always compose two valid parse trees $t_l$ and $t_r$ into a parse tree $\binarytree{A}{t_l}{t_r}$ that is valid as well, as long as $A \to \rootof{t_l}~\rootof{t_r} \in P$.

Based on this observation, we find the following property is crucial to make the $\Rightarrow$-part provable:
any valid parse tree $t$ shall be expanded into a larger valid parse tree $T$ having $t$ as its subtree.
Formally, we define a relation over \emph{signatures} -- each is a pair of nonterminal and sentence:
\begin{definition}(Sub-derivation)
    Let $(A, w)$ and $(B, v)$ be two signatures.
    We say $(B, v)$ is a \emph{sub-derivation} of $(A, w)$, if for every parse tree $t \witness B \derive v$,
    there exists a parse tree $T$ such that $T \witness A \derive w$ and $t$ is a subtree of $T$.
\end{definition}

With this notion, the idea of proving the $\Rightarrow$-part is as follows:
Given $t_1 \neq t_2 \witness A \derive w_A$, if some premise indicates that $(A, w_A)$ is a sub-derivation of $(S, w_S)$ for some $w_S$,
then we can conclude that $S \derive w_S$ is ambiguous by expanding $t_1$ (resp. $t_2$) into $T_1$ (resp. $T_2$), where $T_1 \neq T_2 \witness S \derive w_S$.
To execute this idea, we first propose \emph{reachability} as the missing premise that implies sub-derivation (\cref{sec:ambiguity:reachablity}),
and then define \emph{local ambiguity} to be essentially ``bAMB $\land$ reachability'',
so that ``local ambiguity $\Leftrightarrow$ derivation ambiguity'' becomes provable (\cref{sec:ambiguity:local}).

\subsection{Reachability: The Missing Piece}\label{sec:ambiguity:reachablity}

The definition of reachability should encode necessary conditions that establish a sub-derivation relation.
To make $(B, v)$ a sub-derivation of $(A, w)$, by definition we must prove $A \derive w$ given $B \derive v$,
which can be done via split $A \derive w$ into two big steps: $A \derive \alpha B \beta \derive w_\alpha v w_\beta = w$.
Given $B \derive v$, it suffices to find clauses $\alpha \derive w_\alpha$ and $\beta \derive w_\beta$ such that $A \derive \alpha B \beta$.
Recall that in CFGs, $A \derive \alpha B \beta$ actually means ``$B$ is reachable from $A$''.
This motivates us to extend the reachability relation on CFGs to layout-sensitive grammars -- taking sentences and the layout constraints they should fulfill into account.

\begin{definition}(Reachability)\label{def:reachability}
    The \emph{reachability} relation $(A, w) \reachable (B, v)$ is the reflexive transitive closure of a one-step reachability relation $\reachableonce$ inductively defined as follows:
    \begin{enumerate}[(1)]
        \item If $A \to B^\varphi \in P$ and $\varphi(w)$, then $(A, w) \reachableonce (B, w)$.
        \item If $A \to B \mathbin\varphi B' \in P$, $\varphi(w_1, w_2)$ and $B' \derive w_2$,
            then $(A, w_1 w_2) \reachableonce (B, w_1)$.
        \item If $A \to B' \mathbin\varphi B \in P$, $\varphi(w_1, w_2)$ and $B' \derive w_1$,
            then $(A, w_1 w_2) \reachableonce (B, w_2)$.
    \end{enumerate}
\end{definition}

The three rules for $\reachableonce$ enumerate all possibilities we can derive a nonterminal $B$ from $A$ in one derivation step, via the production rule $A \to B^\varphi$, $A \to B \mathbin\varphi B'$ and $A \to B' \mathbin\varphi B$ respectively.
This definition is expressive enough to indicate sub-derivation relation, as stated in the following lemma
(All lemmas/theorems in this and the next section are mechanized in Coq; we only provide proof sketch in the paper):
\begin{lemma}\label{thm:reachable-sub-derivation}
    If $B \derive v$, then $(A, w) \reachable (B, v)$ iff $(B, v)$ is a sub-derivation of $(A, w)$.
\end{lemma}
\begin{proof}[Proof Sketch]
    For the $\Rightarrow$-part: induction on the reachable relation (in this part, the hypothesis $B \derive v$ is useless).
    For the $\Leftarrow$-part: we have $T \witness A \derive w$ for every $t \witness B \derive v$; induction on $T$ and check in every case $(A, w) \reachable (B, v)$.
\end{proof}

\subsection{Local Ambiguity: Foundation of SMT Encoding}\label{sec:ambiguity:local}

With the reachability relation defined, \emph{local ambiguity} is essentially ``bAMB $\land$ reachability''.
In bAMB, the condition ``two parse trees differ on a node at level 1'' is formally expressed by ``they are not similar'', via a \emph{similarity} relation $\similar$ inductively defined by the rules below:
\begin{mathpar}
    \emptytree{A} \similar \emptytree{A}
    \and \tokentree{A}{tk} \similar \tokentree{A}{tk}
    \and \inference{\rootof{t_1} = \rootof{t_2} & \wordof{t_1} = \wordof{t_2}}
            {\unarytree{A}{t_1} \similar \unarytree{A}{t_2}}
    \and \inference{\rootof{t_{11}} = \rootof{t_{21}} & \wordof{t_{11}} = \wordof{t_{21}}
            & \rootof{t_{12}} = \rootof{t_{22}} & \wordof{t_{12}} = \wordof{t_{22}}}
            {\binarytree{A}{t_{11}}{t_{12}} \similar \binarytree{A}{t_{21}}{t_{22}}}
\end{mathpar}

\begin{definition}[Local ambiguity]\label{def:local-amb}
    A signature $(A, w)$ is said \emph{local ambiguous} if there exists another signature $(H, h)$ such that
    $(A, w) \reachable (H, h)$, and there exist $t_1 \not\similar t_2$ that both witness $H \derive h$.
\end{definition}

Local ambiguity is easier to be encoded than derivation ambiguity: both the similarity and reachability relations are inductively defined, and all side premises are encodable in SMT theories.
Our SMT encoding is essentially translating the local ambiguity definition into a SMT formula, which we will explain in \cref{sec:encoding}.
The following theorem -- ``local ambiguity $\Leftrightarrow$ derivation ambiguity'' -- provides correctness guarantees towards a sound and complete SMT encoding:
\begin{theorem}\label{thm:local_amb_iff}
    $(A, w)$ is local ambiguous iff the derivation $A \derive w$ is ambiguous.
\end{theorem}
\begin{proof}[Proof Sketch]
    For the $\Rightarrow$-part:
    Let $(H, h)$ be a signature such that $(A, w) \reachable (H, h)$.
    Let $t_1 \not\similar t_2$ be two parse trees that both witness $H \derive h$.
    By \cref{thm:reachable-sub-derivation}, there exist $t \witness A \derive w$ and $t_1$ being a subtree of $t$.
    By substituting $t_2$ for $t_1$ in $t$, we obtain a new parse tree $t' \witness A \derive w$.
    We have $t_1 \not\similar t_2 \Rightarrow t_1 \neq t_2 \Rightarrow t \neq t'$.
    Therefore, $t$ and $t'$ witness the ambiguity of $A \derive w$.

    For the $\Leftarrow$-part:
    Let $t_1 \neq t_2 \witness A \derive w$.
    It suffices to extract two subtrees, say $t_1'$ from $t_1$ and $t_2'$ from $t_2$, such that they have the same signature, say $(H, h)$, and that $t_1' \not\similar t_2'$.
    The subtrees can be found via tree difference: compare node pairs between $t_1$ and $t_2$ in a depth-first order and return immediately when the nodes are different.
\end{proof}

By this theorem, we argue the ambiguity of $G_S$ shown in \cref{fig:counterexample}:
Let $w_{ab} = [a@(1,3), b@(2,3)]$, we have:
(1) $(S, w_{ab}) \reachable (A, [a@(1,3)])$,
(2) $t_1 \not\similar t_2$ since $C \neq C'$,
and (3) both $t_1$ and $t_2$ witness $A \derive [a @ (1,3)]$.

\paragraph{Comparison}

Our local ambiguity strengthens bAMB in two dimensions.
First, a reachability relation is defined for layout-sensitive grammars, as a nontrivial extension of the usual reachability notion on CFGs. 
This relation ``locally'' encodes the necessary conditions for ensuring the equivalence theorem on each signature.
Second, our equivalence theorem (\cref{thm:local_amb_iff}) is held for any layout-sensitive grammar, and obviously also for any CFG,
which makes it more general than bAMB that applies to only reduced CFGs.

\section{Bounded Ambiguity Checking}\label{sec:encoding}

With the local ambiguity defined in the previous section, we are now ready to construct an encoding for $\fAmb{k}$, which states that there exists a $k$-length ambiguous sentence, by translating local ambiguity into a SMT formula.
We will present the technical details of this translation in a top-down manner (\cref{sec:encoding:model}, \cref{sec:encoding:encoding})
and show that it is sound and complete (\cref{sec:encoding:correctness}).
Relying on a SMT solver (\eg Z3) as the backend, our bounded ambiguity checker is facilitated by an incremental loop which finds the lowest $k > 0$ such that $\fAmb{k}$ is satisfiable, whose satisfying model gives a shortest ambiguous sentence of the input grammar (\cref{sec:encoding:loop}).

\subsection{Satisfying Model}\label{sec:encoding:model}

Before presenting $\fAmb{k}$, let us see which variables should be included in the satisfying model of this formula.
Above all, we encode the ambiguous sentence $w$ -- a $k$-length positioned token sequence --
by three groups of variables $\mTerm{i}$, $\mLine{i}$ and $\mCol{i}$ (for $0 \le i < k$):
they resp. encode the terminal, the line number and the column number of each positioned token in the sequence.

Besides, we introduce auxiliary propositional variables to state that whether a derivation or reachability relation holds, as required by the definition of local ambiguity.
The variables are split into three groups (where $w_{x, \delta}$ denote the $\delta$-length subword of $w$ starting at index $x$):
\begin{enumerate}
    \item $\mDerive{A}{x}{\delta}$ for $A \in N, 0 < \delta < k - x$ states whether $A \derive w_{x, \delta}$;
    \item $\mReachableNull{A}$ for $A \in N$ states whether $(S, w) \reachable (A, \emptystr)$;
    \item $\mReachable{A}{x}{\delta}$ for $A \in N, 0 < \delta < k - x$ states whether $(S, w) \reachable (A, w_{x, \delta})$.
\end{enumerate}
We use two groups of variables to encode the reachability relation, $\mReachableNull{A}$ for the empty subword and $\mReachable{A}{x}{\delta}$ for nonempty subword $w_{x, \delta}$.
For the derivation relation, however, we do not need variables $\mDeriveNull{A}$ to encode $A \derive \emptystr$, because nullability can be precomputed.
We write $\nullable{A}$ to denote that $A$ is nullable.

\subsection{SMT Encoding}\label{sec:encoding:encoding}

We now present the top-level encoding for $\fAmb{k}$, where the auxiliary definitions $\fDerive{k}$, $\fReachableNull{k}$, $\fReachable{k}$ and $\fMulti{H}{x}{\delta}$ will be introduced later:
\begin{equation*}
\fAmb{k} := \fDerive{k} \land \fReachableNull{k} \land \fReachable{k}
    \land \bigvee_{H \in N} \left(
        (\mReachableNull{H} \land \fMulti{H}{0}{0}) \lor
        \bigvee_{0 < \delta \le k - x} (\mReachable{H}{x}{\delta} \land \fMulti{H}{x}{\delta})
    \right).
\end{equation*}

The first three conjuncts $\fDerive{k}$, $\fReachableNull{k}$ and $\fReachable{k}$ resp. provide logical restrictions on $\mDerive{A}{x}{\delta}$, $\mReachableNull{A}$ and $\mReachable{A}{x}{\delta}$, so that when $\fAmb{k}$ is satisfiable, their truth values correspond to the validity of the derivation/reachability relation as mentioned in \cref{sec:encoding:model}.

The last conjunct encodes local ambiguity by \cref{def:local-amb}:
there exists a nonterminal $H$ and a subword $w^m_{x, \delta}$ such that
(1) $(S, w^m) \reachable (H, w^m_{x, \delta})$, expressed by $\mReachableNull{H}$ (when $w^m_{x, \delta} = \emptystr$) 
or $\mReachable{H}{x}{\delta}$ (when $w^m_{x, \delta} \neq \emptystr$), and
(2) there are two dissimilar parse trees that witness $H \derive w^m_{x, \delta}$, expressed by $\fMulti{H}{x}{\delta}$.

\paragraph{Well-foundedness}

Realizing that derivation and reachability relations are recursively defined on nonterminals, the nonterminal set $N$ should be \emph{well-founded} so that \emph{sound} encodings for $\fDerive{k}$, $\fReachableNull{k}$, $\fReachable{k}$ and $\fMulti{H}{x}{\delta}$ can be constructed.
To obtain a well-founded relation on $N$, we require the grammar to be \emph{acyclic}, which intuitively means any cyclic derivation such as $A \deriveplus A$ is not allowed.
This requirement does not weaken the practicality of our approach: a well-designed grammar should never be cyclic.

Formally, a LS2NF $(N,\Sigma,P,S)$ is said \emph{acyclic} if its graph representation $\langle N, E \rangle$ is acyclic (in graph theory), where $(A, B) \in E$ is an directed edge if
\begin{equation*}
    (A \to B^\varphi \in P) \lor
    (A \to B \mathbin\varphi B' \in P \land \nullable{B'}) \lor
    (A \to B' \mathbin\varphi B \in P \land \nullable{B'}).
\end{equation*}
Checking the acyclicity of a directed graph is solvable in linear-time \cite{tarjan1972depth}.
In the preprocessing phase of \ourtool, if cycles are detected, they will be reported to the user so that the user can manually eliminate them.
In an acyclic grammar, the edge set $E$ of its graph representation is a well-founded relation, called \emph{predecessor} relation, written $A \prec B$.
The inverse (or transpose) relation $E^{-1}$ is also well-founded, called \emph{successor} relation, written $A \succ B$.

\paragraph{Encoding Derivation}

The key observation is that the derivation relation $A \derive w$ ($w \neq \emptystr$) is logically equivalent to the following ``more verbose'' disjunctive form:
\begin{alignedeq}\label{eq:derive-disj}
    (A \derive w) \lequiv &\bigvee_{A \to a \in P} (|w| = 1 \land \Term{w_i} = a)
        \lor \bigvee_{A \to B^\varphi \in P} (B \derive w \land \varphi(w)) \\
        &\lor \bigvee_{\substack{A \to B_1 \mathbin\varphi B_2 \in P \\ w = w_1 w_2}}
            (B_1 \derive w_1 \land B_2 \derive w_2 \land \varphi(w_1, w_2))
\end{alignedeq}
where the three disjuncts enumerates all possible ways to achieve $A \derive w$ by production rule $A \to \alpha \in P$: $\alpha$ can be an atomic, a unary or a binary clause.
Formula $\fDerive{k}$ provides restrictions that every $\mDerive{A}{x}{\delta}$ is true iff $A \derive \decode{m}{x}{\delta}$ -- the latter is encoded as the above disjunctive form:
\begin{align*}
    \fDerive{k} &:= \forall A \in N, 0 < \delta \le k - x: \mDerive{A}{x}{\delta} \lequiv \\
        &\bigvee_{A \to a \in P} (\delta = 1 \land \mTerm{x} = a)
        \lor \bigvee_{A \to B^\varphi \in P} (\mDerive{B}{x}{\delta} \land \fUnaryLayout{\varphi}{x}{\delta})
        \lor \bigvee_{A \to B_1 \mathbin\varphi B_2 \in P} \\
        &\quad\left(
            (\nullable{B_1} \land \mDerive{B_2}{x}{\delta})
            \lor (\nullable{B_2} \land \mDerive{B_1}{x}{\delta})
            \lor \bigvee_{0 < \delta' < \delta}
                (\mDerive{B_1}{x}{\delta'} \land \mDerive{B_2}{x + \delta'}{\delta - \delta'}
                    \land \fBinaryLayout{\varphi}{x}{\delta'}{\delta})
        \right)
\end{align*}
In this formula, the three disjuncts on the rhs of $\lequiv$ respectively correspond to the three disjuncts in \cref{eq:derive-disj}.
The last disjunct of \cref{eq:derive-disj} is split into three cases, depending on if $w_1$ or $w_2$ is empty.
This is necessary because $\mDerive{B_1}{x}{\delta'}$ is only defined for nonempty subwords ($\delta' > 0$).

Moreover, for every possible layout constraint $\varphi$ used in the grammar,
we assume there is a SMT formula that \emph{consistently} encodes its semantics:
(1) for unary constraint $\varphi$, $\Phi_\varphi(x, \delta)$ holds iff $\varphi(w^m_{x, \delta})$;
(2) for binary constraint $\varphi$, $\Phi_\varphi(x, \delta', \delta)$ holds iff
    $\varphi(w^m_{x, \delta'}, w^m_{x + \delta', \delta - \delta'})$.
The encodings for the built-in layout constraints are trivially obtained via a direct translation of their definitions (\cref{fig:layout-predicate}).

\begin{lemma}\label{thm:derivation-encoding}
    If $m \models \fDerive{k}$, then for every $A \in N$, $0 < \delta < k - x$,
    $m(\mDerive{A}{x}{\delta}) = \true$ iff $A \derive w^m_{x, \delta}$.
\end{lemma}
\begin{proof}[Proof Sketch]
    By well-founded induction on $\delta$ and $\prec$.
\end{proof}

\paragraph{Encoding Reachability}

Following the same idea, we represent the reachability relation in a disjunctive form:
\begin{alignedeq}\label{eq:reachable-disj}
    (S, w) \reachable (B, w_B) \lequiv
    \begin{aligned}[t]
        &(B = S \land w_B = w)
        \lor \bigvee_{A \to B^\varphi \in P} ((S, w) \reachable (A, w_B) \land \varphi(w_B)) \\
        &\lor \bigvee_{A \to B \mathbin\varphi B' \in P}
            ((S, w) \reachable (A, w_B w') \land B' \derive w' \land \varphi(w_B, w')) \\
        &\lor \bigvee_{A \to B' \mathbin\varphi B \in P}
            ((S, w) \reachable (A, w' w_B) \land B' \derive w' \land \varphi(w', w_B))
    \end{aligned}
\end{alignedeq}
This form encodes $\reachable$, the reflexive and transitive closure of $\reachableonce$, by
explicitly stating reflexivity as the first disjunct,
and then integrating transitivity into $\reachableonce$, yielding the last three disjuncts that respectively correspond to the three rules for $\reachableonce$ as defined in \cref{def:reachability}.
Depending on whether $w_B$ is empty, we obtain the encodings of $\fReachableNull{k}$ and $\fReachable{k}$:
\begin{align*}
    \fReachableNull{k} &:= \forall B \in N: \mReachableNull{B} \lequiv \\
        &(B = S \land k = 0) \lor
        \bigvee_{A \to B^\varphi \in P} \mReachableNull{A} \lor
        \bigvee_{\substack{A \to B \mathbin\varphi B' \in P \\or A \to B' \mathbin\varphi B \in P}} \left(
            (\mReachableNull{A} \land \nullable{B'}) \lor
            \bigvee_{0 < \delta \le k - x} (\mReachable{A}{x}{\delta} \land \mDerive{B'}{x}{\delta})
        \right) \\
    \fReachable{k} &:= \forall B \in N, 0 < \delta \le k - x: \mReachable{B}{x}{\delta} \lequiv \\
        &(B = S \land x = 0 \land \delta = k)
        \lor \bigvee_{A \to B^\varphi \in P} (\mReachable{A}{x}{\delta} \land \fUnaryLayout{\varphi}{x}{\delta}) \\
        &\lor \bigvee_{A \to B \mathbin\varphi B' \in P}
            (\mReachable{A}{x}{\delta + \delta'}
                \land \ite{\delta' = 0}{\nullable{B'}}{\mDerive{B'}{x + \delta}{\delta'}}
                \land \fBinaryLayout{\varphi}{x}{\delta}{\delta'}
            ) \\
        &\lor \bigvee_{A \to B' \mathbin\varphi B \in P}
            (\mReachable{A}{x - \delta'}{\delta' + \delta}
                \land \ite{\delta' = 0}{\nullable{B'}}{\mDerive{B'}{x - \delta'}{\delta'}}
                \land \fBinaryLayout{\varphi}{x - \delta'}{\delta'}{\delta}
            )
\end{align*}
The layout predicates $\varphi(w_B)$, $\varphi(w_B, w')$ and $\varphi(w', w_B)$ of \cref{eq:reachable-disj} are not translated in $\fReachableNull{k}$ because they trivially hold ($w_B = \emptystr$).
When translating the derivation relation $B' \derive w'$ of \cref{eq:reachable-disj}, we must be careful that $w'$ might be empty:
In $\fReachableNull{k}$, $w'$ is always empty so the translation is $\nullable{B'}$.
In $\fReachable{k}$, we use the ``if-then-else'' predicate $\ite{c}{\Phi_1}{\Phi_2}$, a shorthand for $(c \land \Phi_1) \lor (\lnot c \land \Phi_2)$, to encode both cases.

\begin{lemma}\label{thm:R-enc-empty}
    If $m \models \fDerive{k} \land \fReachableNull{k}$, then for every $B \in N$,
    $m(\mReachableNull{B}) = \true$ iff $(S, w^m) \reachable (B, \emptystr)$.
\end{lemma}
\begin{proof}[Proof Sketch]
    By well-founded induction on $\succ$ (reverse of $\prec$). Apply \cref{thm:derivation-encoding} where necessary.
\end{proof}

\begin{lemma}\label{thm:R-enc-nonempty}
    If $m \models \fDerive{k} \land \fReachable{k}$, then for every $B \in N$, $0 < \delta < k - x$,
    $m(R^B_{x, \delta}) = \true$ iff $(S, w^m) \reachable (B, w^m_{x, \delta})$.
\end{lemma}
\begin{proof}[Proof Sketch]
    By well-founded induction on $k - \delta$ and $\succ$. Apply \cref{thm:derivation-encoding} where necessary.
\end{proof}

\paragraph{Encoding Multiple Parse Trees}

Formula $\fMulti{A}{x}{\delta}$ states that there exist two parse trees $t_1 \not\similar t_2$ such that they both witness $A \derive w^m_{x, \delta}$.
Recall that the intuitive meaning of being dissimilar is that $w^m_{x, \delta}$ is derived from $A$ using two distinct production rules for $A$, or using a binary rule $A \to B_1 \mathbin\varphi B_2$ in two distinct ways,
say there exists two ways of splitting $w^m_{x, \delta} = w_{11} w_{12} = w_{21} w_{22}$ such that both prefixes $w_{11}$ and $w_{21}$ are derivable from $B_1$, and both suffixes $w_{12}$ and $w_{22}$ are derivable from $B_2$.
Syntactically, we represent each possible way of derivation as a \emph{using clause}:
$$
\gamma := \varepsilon \mid a \mid B^\varphi \mid B_1^{\delta'} \mathbin\varphi B_2
$$
where in the last case, a length $\delta' \ge 0$ is annotated to restrict that $B_1$ derives a prefix of this length and $B_2$ the rest.
All possible ways of deriving $w^m_{x, \delta}$ from $A$ are given by the following set:
\begin{align*}
    \Gamma(A, \delta) &:=
        \{\epsilon \mid A \to \varepsilon \in P\}
        \cup \{a \mid A \to a \in P\}
        \cup \{B^\varphi \mid A \to B^\varphi \in P\} \\
        &\cup \{B_1^{\delta'} \mathbin\varphi B_2 \mid A \to B_1 \mathbin\varphi B_2 \in P, \delta' \le \delta\}    
\end{align*}

On the other hand, every using clause semantically states that ``under what condition I can derive $w^m_{x, \delta}$'', formally encoded as SMT formulae:
\begin{align*}
    \sem{\varepsilon}_{x, \delta} &:= \delta = 0 \\
    \sem{a}_{x, \delta} &:= \delta = 1 \land \mTerm{x} = a \\
    \sem{B^\varphi}_{x, \delta} &:= \ite{\delta = 0}{\nullable{B}}
        {\mDerive{B}{x}{\delta} \land \Phi_\varphi(x, \delta))} \\
    \sem{B_1^{\delta'} \mathbin\varphi B_2}_{x, \delta} &:=
        \Phi_\varphi(x, \delta', \delta)
        \land \ite{\delta' = 0}{\nullable{B_1}}{\mDerive{B_1}{x}{\delta'}}
        \land \ite{\delta = \delta'}{\nullable{B_2}}{\mDerive{B_2}{x + \delta'}{\delta - \delta'}}
\end{align*}

Finally, formula $\fMulti{A}{x}{\delta}$ states that, there exists two distinct using clauses from $\Gamma(A, \delta)$ such that both can derive $w^m_{x, \delta}$:
\begin{equation*}
\fMulti{A}{x}{\delta} :=
    \bigvee_{\substack{\gamma_1, \gamma_2 \in \Gamma(A, \delta) \\ \gamma_1 \neq \gamma_2}}
        (\sem{\gamma_1}_{x, \delta} \land \sem{\gamma_2}_{x, \delta})
\end{equation*}

\begin{lemma}\label{thm:multi-enc}
    Let $x + \delta \leq k$ and $m \models \fDerive{k}$, then $m \models \Phi_\text{multi}(A, x, \delta)$ iff
    there exist two dissimilar parse trees that witness $A \derive w^m_{x, \delta}$.
\end{lemma}

\subsection{Formal Properties}\label{sec:encoding:correctness}

For any acyclic LS2NF and any length $k \ge 0$, the proposed SMT encoding is sound and complete:
\begin{theorem}[Soundness]\label{thm:sound}
    If $m \models \fAmb{k}$, then $S \derive w^m$ is ambiguous.
\end{theorem}
\begin{proof}[Proof Sketch]
    By \cref{thm:local_amb_iff} it suffices to show local ambiguity, which is straightforward by applying
    \cref{thm:R-enc-empty,thm:R-enc-nonempty,thm:multi-enc}.
\end{proof}

\begin{theorem}[Completeness]\label{thm:complete}
    If there exists $w$ such that $|w| = k$ and $S \derive w$ is ambiguous,
    then $\fAmb{k}$ is satisfiable.
\end{theorem}
\begin{proof}[Proof Sketch]
    By \cref{thm:local_amb_iff} we have $(S, w) \reachable (H, h)$.
    We pick a model $m$ s.t. 
    \begin{align*}
        w^m &= w \\
        m(\mDerive{A}{x}{\delta}) &\lequiv A \derive w^m_{x, \delta} \\
        m(\mReachableNull{B}) &\lequiv (S, w) \reachable (B, \emptystr) \\
        m(\mReachable{B}{x}{\delta}) &\lequiv (S, w) \reachable (B, w^m_{x, \delta})
    \end{align*}
    By definition, $\fDerive{k}$, $\fReachableNull{k}$ and $\fReachable{k}$ are satisfiable.
    Given that $(S, w) \reachable (H, h)$, $h$ must be a subword of $w$, say $h = w_{x_h, \delta_h}$ for some $x_h$, $\delta_h$.
    By \cref{thm:multi-enc}, $\Phi_\text{multi}(H, x_h, \delta_h)$ is satisfiable.
\end{proof}

For every length $k$, $O(k^2)$ SMT variables are created, and for each variable,
an enumeration loop similar to CYK \cite{younger1967recognition} creates $O(k)$ terms,
where we also use $O(n)$ space to go through the $n$ production rules.
Thus, the space complexity for $\fAmb{k}$ is $O(n \cdot k^3)$.

\subsection{Ambiguous Sentence Generation}\label{sec:encoding:loop}

Our bounded ambiguity checker is facilitated by a bounded loop where the bound is specified by the user.
In the $k$-th iteration (initially $k = 1$), logical constraints for finding an ambiguous sentence with length $k$ are encoded as a SMT formula $\fAmb{k}$.
We rely on a backend SMT solver (e.g. Z3) to check its satisfiability:
if it is satisfiable under some model, say $m \models \fAmb{k}$,
then we are able to decode an ambiguous sentence $w^m$ from $m$, and the loop exits immediately;
otherwise, we try the next interactions until the user-specified loop bound is reached.
In this way, a shortest nonempty ambiguous sentence is obtained.
\section{Layout Constraint Synthesis}\label{sec:synthesis}

In \ourtool, user interaction happens when an ambiguous sentence is found.
The user describes his/her intents on how to disambiguate this sentence by formatting it in distinct ways to accord with the parse trees, as demonstrated in \cref{sec:example}.
Then, the layout constraint synthesizer recommends a set of candidate \emph{layout transformation rules},
each of which transforms a layout-free clause into one with layout constraint.
The user selects a subset to accept and \ourtool applies them to the original grammar, yielding a refined grammar with layout constraints added.
Intuitively, the output grammar is more restricted (``less ambiguous'') than the original one.
The more rounds of interactions being made, the closer the refined grammar will be to the user's expectation.

\subsection{Refinement}\label{sec:synthesis:refinement}

The synthesis algorithm aims to \emph{refine} an input grammar.
This notion is defined as follows:

\begin{definition}[Refinement]\label{def:refinement}
    Let $G=(N,\Sigma,P,S)$ and $G'=(N,\Sigma,P',S)$ by two LS2NFs.
    We say $G'$ is a \emph{refinement} of $G$, if for every production rule $r \in P'$, one of the following holds:
    \begin{enumerate}[(1)]
        \item $r \in P$;
        \item $(A \to B) \in P$ and $r = A \to B^\varphi$ for some $\varphi$;
        \item $(A \to B C) \in P$ and $r = A \to B \mathbin\varphi C$ for some $\varphi$.
    \end{enumerate}
\end{definition}

\begin{proposition}\label{thm:refinement-one-rule}
    Let $G'$ be a refinement of $G$. If $w$ is unambiguous under $G$, then it is also unambiguous under $G'$.
\end{proposition}

This proposition explains the key idea of the refinement relation:
no more ambiguous sentence are introduced in the refined grammar, with several layout constraints are added (case (2) and (3) of \cref{def:refinement}).
However, the refined grammar might be equal to the original one as the relation is \emph{reflexive}.
To avoid this useless case, the user is asked to accept \emph{at least one} candidate synthesized by the algorithm (otherwise the algorithm fails), so that the refined grammar will contain fewer ambiguous sentences.
In this way, our grammar refinement loop is converging.

\subsection{Synthesis Algorithm}\label{sec:synthesis:algorithm}

\begin{algorithm}[t]
    \small
    \DontPrintSemicolon
    \KwIn{grammar $G=(N,\Sigma,P,S)$, user's feedback $F$}
    \KwOut{refined grammar $G'$}
    \ForEach{$r \in P$}{
        \If{$r = A \to B$}{
            $\Psi[r] \gets \{B \leadsto B^\varphi \mid \varphi \in \Phi^\text{unary} \}$
        }
        \ElseIf{$r = A \to B_1 B_2$}{
            $\Psi[r] \gets \{B_1 B_2 \leadsto B_1 \mathbin\varphi B_2 \mid \varphi \in \Phi^\text{binary} \}$
        }
    }
    \ForEach{$w \in F$}{
        $t_w \gets$ parse tree of $w$\;
        \ForEach{$t \in t_w$ in depth-first order}{
            \If{$t = \unarytree{A}{t'}$}{
                \Let $B = \rootof{t'}$, $w = \wordof{t'}$\;
                remove $\{B \leadsto B^\varphi \mid \neg\varphi(w) \}$ from $\Psi[A \to B]$
            }
            \ElseIf{$t = \binarytree{A}{t_1}{t_2}$}{
                \Let $B_i = \rootof{t_i}$, $w_i = \wordof{t_i}$ for $i = 1,2$\;
                remove $\{B_1 B_2 \leadsto B_1 \mathbin\varphi B_2 \mid \neg\varphi(w_1, w_2) \}$
                from $\Psi[A \to B_1 B_2]$
            }
        }
    }
    \lIf{$\forall r:~ \Psi[r] = \emptyset$}{\Return ``inconsistent''}
    \Let $\emptyset \ne \Psi' \subseteq \bigcup_{r \in P} \Psi[r]$ be the user accepted candidates\;
    $G' \gets$ apply $\Psi'$ to $G$\;
    \Return $G'$\;
    \caption{Layout Constraint Synthesis}
    \label{algo:refinement}
\end{algorithm}

Our synthesis algorithm (\cref{algo:refinement}) takes an original grammar with a user feedback as input, and produces a refined grammar (when not fail) according to the user's selection of layout transformation rules.
The user \emph{feedback} $F$ is a set of formatted sentences.
The algorithm builds the parse trees $t_w$ for each $w \in F$ under the original grammar $G$ (line 7).
Based on these parse trees, candidate layout transformation rules are synthesized.
Each \emph{layout transformation rule} is in the form of $\alpha \leadsto \alpha'$, by \cref{def:refinement}, the two clauses $\alpha$ and $\alpha'$ must contain the same nonterminals and the left-hand side must be layout-free.
Otherwise, inconsistent layout constraints are potentially introduced in the refined grammar, which is forbidden in our approach.

Realizing that the set of layout constraints one uses in a concrete grammar is \emph{determined} and \emph{finite},
the synthesis algorithm can be facilitated in an \emph{enumerative} fashion:
every possible layout constraint is attempted, and the ones that are \emph{consistent}, \ie all parse trees are still valid when the layout constraint is added, are included in the candidate set.
By default, our algorithm considers all the built-in layout constraints (\cref{sec:grammar}).

In the algorithm, we use $\Psi[r]$ to maintain the set of all layout transformation rules for $r$ that are consistent with the parse trees we have visited so far.
The complete set of candidates is their union $\bigcup_{r \in P} \Psi[r]$.
Before any parse tree is visited, $\Psi[r]$ is initialized to the full set (line 3 and 5), including all possible unary (and binary) layout constraints $\Phi^\text{unary}$ (and $\Phi^\text{binary}$).
Then, we traverse every subtree $t$ in every parse tree $t_w$ of $w \in F$ (line 6 and 8), and remove the inconsistent ones in the corresponding $\Psi[r]$ (note that $r$ matches the structure of $t$), by checking if the layout constraint is fulfilled on $t$ (line 11 and 14).
When all parse trees get processed, $\Psi[r]$ now contains the set of all consistent rules.
If $\Psi[r]$ is empty for all $r$, then $G$ cannot be refined due to inconsistency (line 15).
Otherwise, the user selects a subset of the candidate transformation rules (line 16),
then the algorithm applies them on $G$ (line 17), yielding a refined grammar $G'$ (line 18).

\section{Coq Mechanization}\label{sec:coq}

All the definitions and theorems presented in \cref{sec:ambiguity} and \cref{sec:encoding} are mechanized in the Coq proof assistant.
Our artifact\footnote{
    \url{https://github.com/lay-it-out/LS2NF-theory}
} consists of 10 Coq files and \about \SI{2}{k} lines of code\footnote{
    Excluding comments and blanks, counted by an open-source Unix tool \t{tokei} (\url{https://github.com/XAMPPRocky/tokei}).
}.
It relies on helpers lemmas (mostly on lists) from a popular library \t{coq-stdpp}\footnote{
    \url{https://gitlab.mpi-sws.org/iris/stdpp}
}.
On a MacBook Pro with Apple M1 chip and \SI{16}{GB} memory, a complete verification takes \about \SI{12}{s}.

In our artifact, there is only one axiom we made:
in the type ``$\text{grammar}~\Sigma~N$'' for LS2NFs, the two type parameters $\Sigma$ and $N$ (they resp. encode the terminal and nonterminal set) have \emph{decidable equality},
that is, any two elements are either equal or not equal (\ie $\forall x, y: \{x = y\} + \{x \neq y\}$, using Coq notions).
This is true in reality: the terminal and nonterminal sets consist of a finite number of user-specified symbols, whose equality can be trivially judged.

\section{Implementation}\label{sec:implementation}

We implemented the proposed interactive grammar design framework as a tool called \ourtool, written in Python.

\ourtool reads a user-input layout-sensitive grammar from an EBNF file.
It first preprocesses the grammar and reduces it to a LS2NF.
Next, nullable nonterminals are precomputed performed via breadth-first search \cite{lange2009cnf},
and cycles are detected from the graph representation of the LS2NF by Tarjan's algorithm \cite{tarjan1972depth}.
If any cycle is found, the cyclic rules will be reported and the user is required to remove them manually.

After the preprocessing phase, the bounded ambiguity checker searches a shortest ambiguous sentence (if any) following the bounded loop explained in \cref{sec:encoding:loop}.
SMT formulae are generated according to \cref{sec:encoding:encoding} and we rely on the PySMT library \cite{gario2015pysmt} to interact with the backend SMT solver Z3.
Once the ambiguous sentence is found, the user is then dropped into a read-eval-print loop, typing commands for reformatting this sentence according to its parse trees and inserting the selecting layout constraints from the ones synthesized by \cref{algo:refinement}.

One noteworthy feature of \ourtool is that it allows users to
customize the start symbol used for ambiguity detection.
If a custom start symbol is used, a partial ambiguous sentence is provided in case of ambiguity.
The user is then required to format the partial sentence instead of a full one.
Interaction is henceforth simplified since one merely considers a subset of the grammar.

\section{Evaluation}\label{sec:evaluation}

We evaluate the performance of \ourtool using three experiments.
The first experiment (\cref{sec:evaluation:fragment}) focuses on 6 representative grammar fragments extracted from three well-known layout-sensitive languages: Python, F\# and Haskell.
The second experiment (\cref{sec:evaluation:yaml}) studies the layout-sensitive subset of YAML's full grammar.
To further demonstrate if our tool scales to a complete grammar, we conducted a third experiment (\cref{sec:evaluation:sass}) on the full SASS grammar.

We address the following research questions:
\begin{itemize}
    \item RQ1: How \emph{effective} is \ourtool in eliminating ambiguity of grammars?
    \item RQ2: How \emph{efficient} is \ourtool in reducing the user's burden of designing a layout-sensitive grammar?
    \item RQ3: How \emph{practical} is \ourtool for a complete grammar design?
\end{itemize}
We study RQ1 and RQ2 in the first two experiments, and concentrate on RQ3 in the third experiment.
Source code and data are hosted on Github: \url{https://github.com/lay-it-out/lay-it-out}.

\paragraph{Environment}

All experiments were conducted on a machine with AMD(R) EPYC(TM) 7H12 CPU and
1024 GB memory, running Ubuntu 20.04 and Python version 3.9.7.

\subsection{Small-scale Case Study -- 6 Grammar Fragments}\label{sec:evaluation:fragment}

We extracted 6 representative layout-sensitive grammar fragments from
the language references/manuals of Python, F\# and Haskell.
Eliminating their layout constraints, we obtained 6 layout-free grammars as the initial grammars, displayed in \cref{tb:eval}.
All of them are ambiguous.
Then the initial grammars were fed to \ourtool, and we interacted with it to disambiguate them.
When an ambiguous sentence was presented to us, we attempted to provide a feedback reflecting our understanding of the design choices made in the original language.

\begin{table*}
    \caption{Evaluation results of 6 representative grammar fragments.}
    \label{tb:eval}


\footnotesize
\begingroup
\addtolength{\jot}{-.5em}
\begin{tabular}{p{.015\textwidth}<{\centering}p{.055\textwidth}<{\centering}p{.3\textwidth}<{\centering}p{.14\textwidth}<{\centering}p{.3
\textwidth}<{\centering}}
    \toprule
    \# & Lang. & Initial grammar & Ambiguous word & Refined grammar \\
    \midrule
    1.1 & Python & $\begin{aligned}
        \nt{document} &\to \nt{stmt}^{+} \\
        \nt{stmt} &\to \nt{control-if} \mid \token{pass} \\
        \nt{control-if} &\to (\nt{b-if} ~ \nt{b-elif}^{*}) ~ \nt{b-else}^{?} \\
        \nt{b-if} &\to \token{if} ~ \token{e} ~ \token{:} ~ \nt{stmt}^{+}\\
        \nt{b-elif} &\to \token{elif} ~ \token{e} ~ \token{:} ~ \nt{stmt}^{+}\\
        \nt{b-else} &\to \token{else} ~ \token{:} ~ \nt{stmt}^{+}
    \end{aligned}$
    & \multicolumn{1}{c}{\begin{tabular}[c]{@{}c@{}}\t{if e : pass pass} \\ ($k = 5$)\end{tabular}}
    & \begin{minipage}{0.3\textwidth}
    \emph{Refined rules:}\\
        \[
            \begin{aligned}
                \nt{document} &\to \alignplus{\nt{stmt}} \\
                \nt{b-if} &\to \offside{({\token{if} ~ \token{e} ~ \token{:}} \indents \alignplus{\nt{stmt}})}
            \end{aligned}
        \]
    \end{minipage}\\
    \midrule
    1.2 & Python & Refined grammar of 1.1
    & \multicolumn{1}{c}{\begin{tabular}[c]{@{}c@{}} $\begin{aligned}
        &\t{if}~\t{e}~\t{:}\\
        &\qquad\t{pass}\\
        &\t{else}~\t{:}~\t{pass}\\
        &\t{pass}
    \end{aligned}$ \\ ($k = 8$)\end{tabular}}
    & \begin{minipage}{0.3\textwidth}
    \emph{Refined rules:}\\
        \[
            \begin{aligned}
                \nt{control-if} &\to (\nt{b-if} ~ \nt{b-elif}^{*}) \alignto \nt{b-else}^{?} \\
                \nt{b-else} &\to \offside{({\token{else} ~ \token{:}} \indents \alignplus{\nt{stmt}})}
            \end{aligned}
        \]
    \end{minipage}\\
    \midrule
    1.3 & Python & Refined grammar of 1.2
    & \multicolumn{1}{c}{\begin{tabular}[c]{@{}c@{}} $\begin{aligned}
        &\t{if}~\t{e}~\t{:}\\
        &\qquad\t{pass}\\
        &\t{elif}~\t{e}~\t{:}~\t{pass}\\
        &\t{pass}
    \end{aligned}$ \\ ($k = 9$)\end{tabular}}
    & \begin{minipage}{0.3\textwidth}
    \emph{Refined rules:}\\
    \[
        \begin{aligned}
            \nt{control-if} &\to (\nt{b-if} \alignto \nt{b-elif}^{*}) \alignto \nt{b-else}^{?} \\
            \nt{b-elif} &\to \offside{({\token{elif} ~ \token{e} ~ \token{:}} \indents \alignplus{\nt{stmt}})}
        \end{aligned}
    \]
    \end{minipage}\\
    \midrule
    1.4 & Python & Refined grammar of 1.3
    & \multicolumn{1}{c}{\begin{tabular}[c]{@{}c@{}}$\begin{aligned}
        &\t{if}~\t{e}~\t{:}\\
        &\qquad\t{pass}\\
        &\t{elif}~\t{e}~\t{:}\\
        &\qquad\t{pass}\\
        &\t{elif}~\t{e}~\t{:}\\
        &\qquad\t{pass}\\
        &\t{pass}
    \end{aligned}$ \\ ($k = 13$)\end{tabular}}
    & \begin{minipage}{0.3\textwidth}
    \emph{Refined rules:}\\
    \[
        \begin{aligned}
            \nt{control-if} &\to (\nt{b-if} \alignto \alignstar{\nt{b-elif}}) \alignto \nt{b-else}^{?} \\
        \end{aligned}
    \]
    \end{minipage}\\
    \midrule
    2 & Python & $\begin{aligned}
        \nt{block} &\to \nt{stmt}^{+} \\
        \nt{stmt} &\to \nt{while-stmt} \mid \token{pass} \\
        \nt{while-stmt} &\to \nt{while-test} ~ \nt{block} \\
        \nt{while-test} &\to \token{while} ~ \token{e} ~ \token{:}
    \end{aligned}$
    & \multicolumn{1}{c}{\begin{tabular}[c]{@{}c@{}}\t{while e : pass pass} \\ ($k = 5$)\end{tabular}}
    & $\begin{aligned}
        \nt{block} &\to \alignplus{\nt{stmt}} \\
        \nt{stmt} &\to \offside{(\nt{while-stmt} \mid \token{pass})} \\
        \nt{while-stmt} &\to \offside{(\nt{while-test} \indents \nt{block})} \\
        \nt{while-test} &\to \token{while} ~ \token{e} ~ \token{:}
    \end{aligned}$\\
    \midrule
    3 & F\# & $\begin{aligned}
        \nt{start} &\to \nt{match-id} ~ \nt{rules} \\
        \nt{match-id} &\to \token{match} ~ \token{id} ~ \token{with} \\
        \nt{rules} &\to (\token{|} ~ \token{id} ~ \token{->} ~ (\nt{start} \mid \token{id}))^{+}
    \end{aligned}$
    & \multicolumn{1}{c}{\begin{tabular}[c]{@{}c@{}}\t{match id with | id ->}\\\t{match id with | id}\\\t{-> id | id -> id} \\ ($k = 13$)\end{tabular}}
    & $\begin{aligned}
        \nt{start} &\to \offside{(\nt{match-id} \indents \nt{rules})} \\
        \nt{match-id} &\to \token{match} ~ \token{id} ~ \token{with} \\
        \nt{rules} &\to \alignplus{\offside{(\token{|} ~ \token{id} ~ \token{->} ~ (\nt{start} \mid \token{id}))}}
    \end{aligned}$\\
    \midrule
    4 & F\# & $\begin{aligned}
        \nt{start} &\to \nt{expr}^{+} \\
        \nt{expr} &\to (\nt{l-expr} \mid \nt{m-expr})^{+} \mid \token{e} \\
        \nt{l-expr} &\to \nt{bind} ~ \nt{expr} \\
        \nt{bind} &\to \token{let} ~ \token{id} ~ \token{=} ~ \nt{expr} \\
        \nt{m-expr} &\to \nt{m-with} ~ \nt{rules} \\
        \nt{m-with} &\to \token{match} ~ \token{id} ~ \token{with} \\
        \nt{rules} &\to (\token{|} ~ \token{id} ~ \token{->} ~ \nt{expr})^{+}
    \end{aligned}$
    & \multicolumn{1}{c}{\begin{tabular}[c]{@{}c@{}}\t{match id with}\\\t{| id -> match id with}\\\t{| id -> e}\\\t{| id -> e} \\ ($k = 7$)\end{tabular}}
    & $\begin{aligned}
        \nt{start} &\to \alignplus{\nt{expr}}\\
        \nt{expr} &\to \nt{l-expr} \mid \nt{m-expr} \mid \token{e} \\
        \nt{l-expr} &\to \nt{bind} \alignto \nt{expr} \\
        \nt{bind} &\to \offside{(\token{let} ~ \token{id} ~ \token{=} ~ \nt{expr})} \\
        \nt{m-expr} &\to \offside{(\nt{m-with} ~ \nt{rules})} \\
        \nt{m-with} &\to \token{match} ~ \token{id} ~ \token{with} \\
        \nt{rules} &\to \alignplus{(\offside{\token{|} ~ \token{id} ~ \token{->} ~ \nt{expr})}}
    \end{aligned}$\\
    \midrule
    5 & Haskell & $\begin{aligned}
        \nt{let-expr} &\to \token{let} ~ \nt{decl}^{*} ~ \token{in} ~ \nt{expr} \\
        \nt{decl} &\to \nt{expr} ~ \token{=} ~ \nt{expr} \\
        \nt{expr} &\to \token{id} \mid \token{id} ~ \token{id} \mid \nt{let-expr}
    \end{aligned}$
    & \multicolumn{1}{c}{\begin{tabular}[c]{@{}c@{}}\t{let id = id id}\\\t{id = id in id} \\ ($k = 10$)\end{tabular}}
    & $\begin{aligned}
        \nt{let-expr} &\to \token{let} ~ \alignstar{\nt{decl}} ~ \token{in} ~ \nt{expr} \\
        \nt{decl} &\to \nt{expr} ~ \token{=} ~ \nt{expr} \\
        \nt{expr} &\to \token{id} \mid \token{id} ~ \token{id} \mid \nt{let-expr}
    \end{aligned}$\\
    \midrule
    6 & Haskell & $\begin{aligned}
        \nt{where-clause} &\to \nt{decl} ~ \token{where} ~ \nt{decl}^{+} \\
        \nt{decl} &\to \nt{expr} ~ \token{=} ~ \nt{expr} \\
        \nt{expr} &\to \token{id} \mid \token{id} ~ \token{id}
    \end{aligned}$
    & \multicolumn{1}{c}{\begin{tabular}[c]{@{}c@{}}\t{id = id where id =}\\\t{id id id = id} \\ ($k = 11$)\end{tabular}}
    & $\begin{aligned}
        \nt{where-clause} &\to \nt{decl} ~ \token{where} ~ \alignplus{\nt{decl}} \\
        \nt{decl} &\to \nt{expr} ~ \token{=} ~ \nt{expr} \\
        \nt{expr} &\to \token{id} \mid \token{id} ~ \token{id}
    \end{aligned}$\\
    \bottomrule
\end{tabular}

\endgroup
\end{table*}

\paragraph{Answer to RQ1}

After the interaction rounds presented in \cref{tb:eval} were completed, an additional round was launched for each grammar to check bounded ambiguity with bound 20 -- no ambiguous sentences were found within the length 20.
We also manually inspected the refined grammars, and didn't identify any ambiguous sentences.

\paragraph{Answer to RQ2}

\cref{tb:eval} presents the generated ambiguous sentence (where $k$ indicates its length) and the refined grammar after synthesis.
Our tool makes it easier for grammar designers to detect ambiguity.
For example, in round \#1.4, one author once manually constructed an ambiguous sentence of length 17 with nested if-elif-blocks, and the parse trees are quite deep.
However, as shown in \cref{tb:eval}, \ourtool generated a shorter sentence of length 13, without nested structures, which reduces the user's burden on the analysis of ambiguity.
Four interaction rounds (\#1.1 -- \#1.4) were taken on the Python if-else case; all other five cases took a single interaction round.
In all rounds, the generated ambiguous sentence had two parse trees.

\paragraph{A notable case}

We now discuss the Python if-else case as it took multiple rounds (\#1.1 -- \#1.4).
This grammar fragment is highly ambiguous and finally 13 layout constraints were added to disambiguate.
In other layout-sensitive languages like SASS and Scala 3, a similar fragment to this case are seen.
We now explain the synthesized layout constraints:
(1) all \nt{b-elif} branches must be aligned, same for statements in \nt{document} (\#1.1,\#1.4);
(2) \nt{b-elif} and \nt{b-else} branches must align to \nt{b-if} (\#1.2, \#1.3);
(3.a) \nt{b-*} blocks must obey the offside rule, and statements inside those blocks must be aligned;
(3.b) \nt{if e:} should be laid on the same line, same for \nt{elif e:} and \nt{else:};
(3.c) statements in \nt{b-*} blocks must be indented from the preceding tokens (\#1.1-\#1.3).

\subsection{Middle-Scale Case Study -- YAML's Subset}\label{sec:evaluation:yaml}

YAML is a human-readable serialization format whose grammar is layout-sensitive.
Despite a non-large size (21 rules in EBNF and 37 rules in LS2NF),
YAML covers all types of layout rules in its final grammar,
exhibiting even more versatile ambiguity cases than SASS,
though much smaller than the latter.
In this case study, we concentrate on its layout-sensitive subset.

\paragraph{Answer to RQ1}

Like in the first experiment, we launched an additional round to check bounded ambiguity after the main rounds.
No ambiguous sentences were found within the length 20.

\paragraph{Answer to RQ2}

Considering the complexity of YAML, it is rather difficult to construct ambiguous sentences manually.
However, with \ourtool, the grammar designer is guided to disambiguate it within a few rounds.
The interaction took 3 rounds in total, and the execution time (of formula generation and solving) in each round is under 2 seconds.
The generated ambiguous words were short enough, among them, the longest one was ``\t{?}~\t{t}~\t{:}~\t{t}~\t{:}~\t{t}''
(length 6).
In the final grammar, 8 layout constraints were added.

\subsection{Large-scale Case Study -- SASS}\label{sec:evaluation:sass}

SASS is a CSS (Cascading Style Sheets) preprocessor language widely used in web frontend development.
It is self-described as ``Syntactically Awesome Style Sheets''.
A survey \cite{coyier2012} shows that among all developers who use a CSS preprocessor, SASS took 41\% of the market, being the second most popular option.
As a preprocessor language, SASS has a richer feature set than CSS, such as nested blocks, mixins and control structures.
These rich features use layout-sensitive syntax, which makes SASS a complex and interesting case to measure the practicality of \ourtool on designing a complete grammar (RQ3).

This case study was conducted on the \emph{full grammar} of SASS.
Again, we use the layout-free version as the initial grammar.
The initial grammar contains 564 production rules in LS2NF.
Handling such a large grammar can be challenging, and provides us adequate data so that an analysis of interesting metrics is possible.
Overall, this large case took 16 rounds of interactions and finally 74 layout constraints were introduced.
Each generated ambiguous sentence has 2 parse trees.

\paragraph{Analysis on collected metrics.}

\begin{figure}
    \scalebox{0.5}{\input{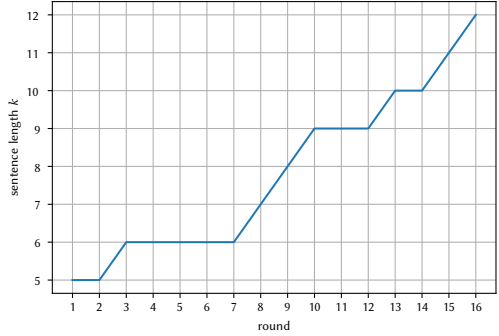}}
    \caption{Ambiguous sentence length in each round.}
    \label{fig:ambig-len}
\end{figure}

The length of the ambiguous sentences found in each round is depicted in \cref{fig:ambig-len}, with 12 being the longest.
Note that in the last round, to exploit the \emph{layered structure} of SASS, 
we changed the start symbol and instead generated an ambiguous subword.

To characterize the complexity of the encoded formula, we look into
the average formula size\footnote{
    The formula size is the number of nodes on its AST representation, computed by the PySMT \cite{gario2015pysmt} library function \lstinline|pysmt.shortcuts.get_formula_size(f)|.
} of different sentence lengths.
Note that SMT solver was invoked many times in our experiment,
and we collected the formula size of each invocation (no matter the solving result is sat or unsat) as samples.
The relationship between SMT formula size and sentence length is plotted in \cref{fig:ambig-node}.
We used linear regression to check if the growth conforms to our theoretical complexity.
The $R^2$ is a statistical measure between -1 and 1, indicating how much one variable relates to another.
The regression line has $R^2 \approx 0.9975$.
Thus, the results are consistent to the theoretical complexity $O(n \cdot k^3)$.

\begin{figure*}
    \centering
    \begin{minipage}[b]{.48\textwidth}
        \scalebox{0.48}{\input{fig/node_vs_len.pgf}}
        \caption{Growth of SMT formula size.}
        \label{fig:ambig-node}
    \end{minipage}
    \vspace{0pt}
    \begin{minipage}[b]{.48\textwidth}
        \scalebox{0.48}{\input{fig/time_vs_len.pgf}}
        \caption{Execution time for encoding and solving $\fAmb{k}$.}
        \label{fig:sass-total-time}
    \end{minipage}
\end{figure*}

We also plotted \cref{fig:sass-total-time} to analyze the execution time under different sentence length $k$.
The total time consists of formula generation time and SMT solving time.
From this figure we see that most iterations finished in \SI{0.5}{h} ($k < 10$),
with an exception of 4 cases.
For cases within \SI{0.5}{h}, the average formula size is 303,695.
In contrast, the SMT formula sizes of other 4 cases, annotated on the figure, are all above ${10}^6$.
Formulae of such size are in general quite hard for SMT solvers.

\paragraph{Answer to RQ3}

The 16 rounds of interaction with \ourtool shows that our approach is feasible and robust for a complete grammar design phase.
The generated ambiguous sentences are relatively short and they are useful to help one detect the cause of ambiguity without much efforts.
Without them, at least for ourselves, it will be impossible to eventually identify all these 74 layout constraints.
Although the SMT solving is not fast on the last 2 rounds, the execution time is still reasonable for a grammar of such a large scale.

\subsection{Discussion}

In our current revision, the bounded ambiguity checker starts from $k = 1$ in every round.
In fact, this is unnecessary.
If $k$ is the length of the found ambiguous sentence of the last, then we can just start from $k$ (instead of $1$) in the next round (as long as the start symbol is unchanged).
This makes sense according to \cref{def:refinement}.

Many real grammars can be naturally split into multiple layers (e.g. \nt{document}, \nt{stmt}, \nt{expr}).
Thus, we allow the user to specify an arbitrary nonterminal as the start symbol for this round.
Then, the user can focus on the ambiguity of this fragment.

For scalability, \cref{sec:evaluation:sass} has already shown that our tool is capable to handle a full grammar.
However, \ourtool is limited by the efficiency of the backend SMT solver if the formula size is in the millions, such as the last two rounds (\cref{fig:sass-total-time}).

\section{Related Work}\label{sec:related}

\paragraph{Layout-sensitive Languages \& Parsing}

In 1966, \citet{Landin} first introduced layout-sensitive languages, which has influenced the syntax design of many later programming languages.
As an extension to CFG, \emph{Indentation-sensitive CFG} (ISCFG) \cite{Revisiting-Offside} is expressive of layout rules
and derives LR($k$) and GLR algorithms for parsing these layout-sensitive grammars.
In ISCFG, symbols are annotated with the numerical relation that the indentation of every nonterminal must have with that of its children.
Although ISCFG has a formal theory and is parser-friendly,
it takes too much effort for a human to manually specify those annotations.

In a declarative layout specification \cite{Layout-GLR}, layout constraints are expressed with primitives that support direct access to the position of a certain ``border'' of a code block.
To realize layout-sensitive parsing, a naive approach is to first generate every possible parse tree with GLR parsing while neglecting layouts, and then filter with layout constraints.
This is, however, inefficient for practical applications.
To improve performance, they identify a subset of constraints that are independent from context-sensitive information and enforce them at parse time.
Later, another version with high-level specification is proposed by \citet{Decl-Spec}.
Our grammar declarative specification notations are inspired by theirs.
The difference lies in the way we regard layout constraints --
in their specification, constraints are attached to a normal layout-free production rule,
whereas in ours, constraints are part of the rule.
Apart from generalized parsing, \tool{Iguana} \cite{One-Parser,Iguana} is a novel parsing framework based upon \emph{data-dependent grammars} \cite{Data-dependent-Grammars},
which extends a CFG with arbitrary computation, variable binding and logical constraint.
\tool{Iguana} translates high-level declarations into equations that are expressive in data-dependent grammars.

Brzozowski derivatives \cite{Brzozowski} have been rediscovered recently to simplify the explanation to parsers.
\citet{First-class-Derivatives} propose a new parser combinator library with first-class derivatives, gaining fine-grained control over an input stream.
It is still an open question whether their framework can implement alignment and offside rules in a modular way, e.g. the Haskell's grammar.

\paragraph{Word Enumeration}

\citet{Grammar-Comparison} propose a practical approach to check the equivalence of CFGs, based on random enumeration of parse trees and words.
To generate words of a fixed length, they first transform the input CFG into a restricted CFG that can only derive words of the given length, and then apply the random enumeration on it.
Based on our knowledge, it is more challenging to apply random enumeration techniques to generate ambiguous words.
Instead, constraint solving is more suitable to solve such a conditional search problem.

\paragraph{Grammar Synthesis}

Is it possible to generate a parser from examples?
\citet{Parser-Gen-from-Examples} raised this question and attempted with genetic programming methods.
In the recent decade, the research community has significant interest in \emph{programming by examples} \cite{flash-fill,flash-meta}
and novel approaches have been proposed to automate parser construction.

\tool{Parsify} \cite{Parsify} is a graphical, interactive system for synthesizing and testing parsers from user-provided examples (a set of sentences).
They rely on a GLL parser to identify ambiguous grammars and a few \emph{disambiguation filters} \cite{Disambig-Filters,Disambig-Sub-parse} to eliminate the well-known associatively and priority issues in grammars of binary expressions.
They disambiguate in an interactive manner: possible parse trees are presented to the user, and only one of them is accepted.
This does not apply in our situation where all parse trees are acceptable \emph{simultaneously} due to the inadequate layout information.

\tool{Glade} \cite{Glade} is an oracle-based grammar inference system.
The algorithm synthesizes a CFG which encodes the language of valid program inputs, beginning with a small set of the target language, provided by the user, as seed inputs.
Although grammar synthesis is a promising direction that can ease the design of a grammar and parser,
based on our knowledge, there is still no work on the automated synthesis of layout-sensitive grammars and parsers.

\paragraph{Grammar-based Fuzzing}

To improve the test converge on programs whose inputs are highly structured, like compilers and interpreters,
grammar-based fuzzing is proposed to leverage a user-defined grammar for generating syntactically valid inputs.
Black-box fuzzing has been integrated with manually specified grammars to test C compilers \cite{Random-C,Csmith},
find bugs in PHP and JavaScript interpreters \cite{LangFuzz},
and generate plausible inputs with the help of a parser merely \cite{Parser-Fuzz}.
There are also studies integrated with white-box techniques.
For example, \citet{Whitebox-Fuzzing} use a handwritten grammar in combination with a custom grammar-based constraint solver to fuzz a JavaScript interpreter of Internet Explorer 7.
CESE \cite{Symbolic-Grammars} combines exhaustive enumeration of valid inputs with symbolic execution.
In comparison, our problem is to find an ambiguous sentence of the grammar, instead of just a random sentence accepted by the grammar.

\section{Conclusion \& Future Work}

We present \ourtool, a framework for layout-sensitive grammar design via user interaction.
Our SMT-based bounded ambiguity checker produces shortest ambiguous sentences that help a grammar designer to recognize the ambiguity during the design phase.
With a polynomial-space encoding, the checker scales to the SASS full grammar.
Through user interactions, \ourtool recommends candidate layout constraints for the grammar designer to select, which reliefs the user from manually specifying layout rules.
In future, we plan to enhance the prototype tool implementation, such as graphical frontend support for a better visualization of parse trees and a smoother user-experience on interaction.


\bibliography{parsing,synthesis,smt,tool}

\end{document}